\documentstyle[prl,aps,epsfig]{revtex}

\begin{document}

\date{\today}
\draft
\twocolumn

\title{
Hot electron transport in 
Ballistic Electron Emission Spectroscopy:
band structure effects and
k-space currents 
}

\author{
K. Reuter$^\ddagger$, P.L. de Andres 
}

\address{
Instituto de Ciencia de Materiales (CSIC), 
Cantoblanco, E-28049 Madrid (SPAIN)
}

\author{
F.J. Garcia-Vidal, F. Flores
}

\address{
Dept. de Fisica Teorica de la Materia Condensada (UAM), 
Universidad Autonoma de Madrid, E-28049 Madrid (SPAIN)
}

\author{
U. Hohenester, P. Kocevar
}

\address{
Institut f\"ur Theoretische Physik, 
Karl-Franzens- Universit\"at Graz,
A-8010 Graz (AUSTRIA)
}

\maketitle

\begin{abstract}
Using a Green's function approach, we investigate band structure
effects in the BEEM current distribution in reciprocal space. In the
elastic limit, this formalism provides a "parameter free" solution
of the BEEM problem.  At low temperatures, and for thin 
metallic layers, the elastic approximation is enough to explain 
the experimental I(V) curves at low voltages. At higher voltages inelastic
effects are approximately taken into account by introducing an
effective RPA-electron lifetime, much in similarity with 
LEED theory. For thick films, however, additional
damping mechanisms are required to obtain agreement with experiment.

\end{abstract} 


Ballistic Electron Emission Microscopy (BEEM),
and its spectroscopic counterpart (BEES)\cite{kaiser},
were originally designed as techniques extending the power
of Scanning Tunneling Microscopy (STM) to buried
interfaces, particularly of metal-semiconductor systems.
The standard model describes BEEM as a 
convolution of three
steps\cite{mario}: tunneling from the tip (1),
propagation in the metallic layer (2) and 
transmission through the metal-semiconductor
interface (3).
This model clearly suggests 
the important potential of
BEEM to focus in any of these steps separately.
However,
it is
unnecessary to stress that such a deconvolution
process may only be safely performed applying
a sufficiently elaborated theory,
which should use as few adjustable parameters as
possible.
In the past, the lack of such a precise method
to analyze the experiment has prompted several
intense discussions: (i) whether 
$k_{\parallel}$ is conserved or not
at the interface\cite{ludeke}, (ii) the origin of
the observed nanometric resolution and
its relation to the tunneling
injection\cite{milliken}, (iii) the similar results
obtained on
Au/Si(111) and Au/Si(100) interfaces, despite
of their different projected conduction-band minima\cite{schowalter},
(iv) 
how {\it ballistic} are the
electrons in BEEM after all?\cite{mario},
etc.
This list of intensively debated
questions in the literature is probably
an indication of the limitations associated
with the standard approach based on E-space
Monte-Carlo simulations, where
processes crucial from a physical point
of view are simply parametrized
to give agreement with experiment.
In particular, in all these Monte-Carlo 
calculations, the energetic spectrum and the 
momentum distribution of the injected electrons are 
taken from conventional planar tunneling theory,  
using a free electron approach. This assumption 
is probably the origin of the 
major limitation to a first-principles analysis 
of the BEEM current, as the propagation of the 
electrons in the metal film is strongly dependent 
on the metal band structure and can depart significantly 
from a free electron behaviour \cite{prl}. 
Accordingly, the aim of this letter is to present 
a microscopic formalism that incorporates those 
band structure effects and yields the appropriate 
angular momentum distribution that, as shown below for the 
case of gold films,  
drastically departs from the narrow forward 
cone assumed in E-space Monte-Carlo simulations. 
We will show how most of 
the previous interpretations of BEEM data for Au/Si interfaces
need to be modified when using the right k-space currents. 

\begin{figure}
\vskip -1cm
\epsfig{figure=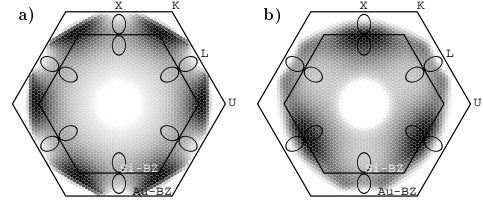,width=8cm}
\vskip -7cm
\caption{
(a) k-space current distribution for $Au(111)$ 
inside the coherence region (5th atomic layer)
($V=1$ eV,
$\eta=0.1$ eV);
(b) same as (a), but outside the coherence
region (30th atomic layer). Dark regions
correspond to higher intensities. The Si
BZ (small) and the Au BZ (large) are shown
together with the ellipses where the
Si conduction band minima project (notice
that the outer ellipses appear after
the corresponding remapping).
}
\end{figure}

We introduce a full quantum-mechanical
description of the BEEM problem based on a Keldysh
Green's function method written in a
Linear Combination of Atomic Orbitals (LCAO)
basis. 
Our analysis is based on 
the following three-step scheme:
We provide an accurate description of the initial tunneling  
injection (1), and the subsequent propagation of electrons
through the metalic layer (2). Passing over the Schottky
barrier (3) is taken into account applying 
energy and $k_{\parallel}$ conservation, and
matching states at the {\it two-dimensional} interface. 
In this paper the foregoing scheme is applied
to the case study of a 
(111) oriented gold metallic layer
deposited on a (111) silicon substrate;
applications to other metals (e.g. $CoSi_{2}$)
and other semiconductors are in progress.
Perfect unrelaxed surfaces and bulk-like 
ideal geometries are assumed in our analysis,
but it is seen from the nature of our results
that a relatively small amount of disorder  
(e.g., confined to 3-4 layers
close to the interface) would not fundamentally
change our conclusions.

A Green's function formalism presents
the important advantage of being free of any
adjustable parameter in the strictly
elastic limit, where we only add an arbitrarily small positive
imaginary part
to the energy ($\eta$), necessary to ensure 
attenuation of the wave at infinity.
Moreover,
inelastic effects associated with the
electron-electron interaction can be added
incorporating a complex energy dependent
self-energy, $\eta(E)$. 
We shall view the self-energy as a single parameter
to be adjusted to the experiment, representing an 
effective inelastic electron-electron mean free path
producing attenuation:
$\lambda_{att} \approx { \sqrt{2E} \over 2 \eta }$.
This method has been 
succesfully adopted to different fields, like
Low-Energy Electron Diffraction.

In an LCAO basis, we write the Hamiltonian
as:

\begin{equation}
\hat H = \hat H_{T} + \hat H_{S} + \hat H_{I}
\end{equation} 

\noindent
where $\hat H_{T} = \sum \epsilon_{\alpha}
\hat n_{\alpha} + \sum \hat T_{\alpha \beta}
\hat c_{\alpha}^{\dagger} \hat c_{\beta}$ defines the
tip (Greek subindices),
$\hat H_{S} = \sum \epsilon_{i}
\hat n_{i} + \sum \hat T_{i j}
\hat c_{i}^{\dagger} \hat c_{j}$ designates the
metal substrate (Latin subindices), and
$\hat H_{I} = 
\sum \hat T_{\alpha j}
\hat c_{\alpha}^{\dagger} \hat c_{j}$ describes the
coupling between the tip and the metal surface
in terms of a
hopping matrix, $\hat T_{\alpha j}$,
expressed as a function
of the different atomic orbitals in the tip and
the surface by using a tight-binding 
formalism\cite{slater,constan} ($\hat n_{\alpha}$,
$\hat c_{\alpha}^{\dagger}$, and $\hat c_{\alpha}$,
are number, creation and destruction operators
defined in the usual way).

Since
the system under investigation is out of
equilibrium, 
a convenient way to compute the current
between two sites i and j in real space
is given by
Keldysh's
technique\cite{keldysh}:

\begin{equation}
J_{i j} = 
\int Tr \{ \hat T_{i j} 
(\hat G_{ij}^{+-} - \hat G_{ji}^{+-}) \} d E 
\end{equation}

\noindent
The matrices $\hat G_{ji}^{+-}$ 
are non-equilibrium Keldysh Green's functions
that can be calculated
in terms of the standard retarded and advanced Green's 
functions\cite{prl,phscr}.
We notice that this formalism allows us to compute 
on the same footing the tunneling current between the tip and the sample
and the current propagating in the metal
(steps 1 and 2).
To this point, all our expressions are exact, and
the main task is to determine
how to compute the
retarded and advanced Green functions, and which
approximations are introduced there. 


Previously\cite{prl}, we have analyzed the 
electron propagation in real space, using a semiclassical 
approximation for these Green functions, and have 
found important focusing effects in gold films. Now, we concentrate 
on calculating the full quantum-mechanical  
current distribution in reciprocal space, using 
a formalism based on renormalization group techniques \cite{guinea}.  
This momentum distribution will allow us to obtain
the spectral I(V) characteristics.
In particular, the current 
between two layers
$a$ and $b$ inside the metal, at
a given energy E and $k_{\parallel}$,
can be expressed as\cite{prl,applied}:

\begin{equation}
J_{ab} (E,k_{\parallel})
= {2 e \over \pi \hbar} \Re \
Tr \{ 
\hat T_{ab}
\hat g_{b1}^{R}
\hat T_{10} \hat \rho_{00} \hat T_{01} 
\hat g_{1a}^{A}
\}
\end{equation}

\noindent
where $\hat g_{b1}^{R(A)}(E,k_{\parallel})$ 
is the retarded (advanced)
Green's function for the unperturbed metal linking the layer $b$ and 
the surface layer, $1$,
$\hat T_{ab}(k_{\parallel})$ is a hopping matrix
connecting layers $a$ and $b$, and
$\hat \rho_{00}(E)$ is the density of states
on the last atom of the tip ($0$),
considered for simplicity to be the
only tip active atom for tunneling.
The trace denotes 
a summation over the orbitals forming the
chosen basis.

Step three of our four-step scheme involves computing the
transmission coefficient for the
two-dimensional interface.
Applying a surface Green's function matching
formalism\cite{verde} in the neighbourhood of the
$\overline{M}$ point,
we obtain a
transmission coefficient
$T(E,k_{\parallel})$
that can be used in
k-space to give the injected
current in the semiconductor:

\begin{equation}
I(V)=
\int_{E_{F}+eV_{0}}^{E_{F}+eV} d E
\int_{1st B.Z.} d k_{\parallel}
J_{c-1,c}(E,k_\parallel) \times T(E,k_{\parallel})
\end{equation}

\noindent
where $c$ refers to the metal layer
at the interface, and $V_{0}$ is
the Schottky barrier height (assumed to be 0.86 eV);
note that the transmission coefficient is zero
outside the 
ellipsoids allowed by energy
conservation (see Figure 1).
The integral inside the first Brillouin
zone is performed
summing over
a dense grid of special points\cite{rafa}.

In previous publications we have discussed
how the propagation of electrons in the gold 
periodic lattice results in
focused beams and narrow Kossel-like lines
in real space,
with a 3-fold
symmetry associated to the $(111)$ direction
of an fcc crystal
\cite{prl,phscr,applied}.
These lines have typical widths of around 3-4
atomic distances, explaining the nanometric
resolution of the BEEM technique even in deeply
buried interfaces. These results also show how the
Bloch wave is formed after propagation
by more than four or five layers, 
forbidding the propagation of electrons
in gap directions over longer distances.
We believe that our results are convincing enough to
answer a question nowadays found in the literature
related to Monte-Carlo simulations:
is it realistic to assume that 
electrons can propagate 
$20$ or $30$ {\AA} as free particles
along the forbidden Au(111) directions? 
We conclude that this is an 
unphysical scenario because of the
strong deflection exerted by the lattice
on electrons
traveling in these directions.

However, in this work we shall focus on 
our results in reciprocal
space and their influence on the I(V) curves. 
An important feature observed
in reciprocal space is a change in the symmetry 
of the k-space current distributions when going 
from thin to thick layers.
The expected symmetry for a
quantum-mechanical calculation is related
to the projected density of states\cite{stiles}. It is
six-fold in (111) fcc planes, because of the
equal contribution of $+\vec k$ and $-\vec k$
states. This is indeed the
case for an arbitrarily small imaginary part
($\eta$)
added to the energy, 
but as commented above, 
$\eta$ 
can be interpreted in terms of a
complex self-energy arising from inelastic
events defining a coherence region
of the order of $\lambda_{att}$. 
Beyond that region
inelastic processes become important, and intensities
rather than amplitudes add to give
the final wavefield. This takes us from a quantum
mechanical picture (six-fold)
to a semi-classical one (three-fold), as 
can be seen comparing Figure 1a,
inside
the coherence region, to Figure 1b,
where the current distribution is computed
in a layer outside that area.
The three-fold symmetry is progressively
built up as a function of metal thickness,
and can be understood in terms of our
previous analysis\cite{prl}: 
the symmetry of the wavefield in the 
semiclassical limit is related to the 
Fermi surface, reflecting the three-fold 
symmetry of the crystal.
Therefore, this is a new example of how
a quantum system,
under the influence of friction,
becomes gradually
classical by a decoherence process\cite{zurek}.
In addition, it is seen how the
current in k-space deviates for
these thick layers from
a simple density of states calculation\cite{stiles},
concentrating around
the directions predicted by the semiclassical
analysis (Figure 1b)\cite{prl,phscr}.
The difference observed in reciprocal space 
between the quantum
and semiclassical regime does not significantly affect
\vskip -4.5cm
\begin{figure}
\epsfig{figure=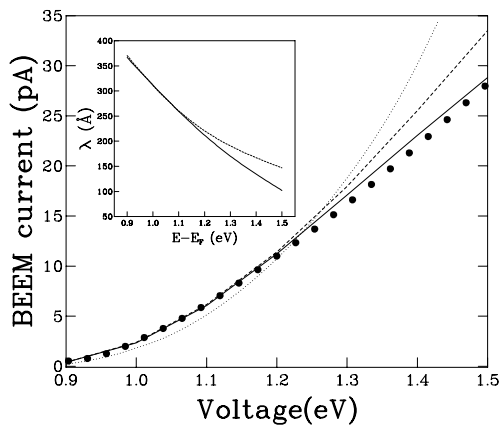,width=8cm}
\caption{
Theoretical I(V) curves for Au/Si(111), 
d=75 \AA\ (experimental values 
-solid circles- from ref. 17): ballistic -dotted-,
RPA approximation for $\lambda_{att}(E)$ (see Eq.5)
with $\lambda^{0}=260$ {\AA}(eV)$^2$ -dashed line- and 
with $\lambda_{att}$ modified -solid line- are shown. In the inset 
these two $\lambda_{att}(E)$ are displayed.}
\end{figure}
the beams in real space (where the symmetry must always
be three-fold), but could in principle affect the
I(V) current injected through the projected
ellipses into the semiconductor. However,
because of the gradual crossover seen from one regime
to the other, we do not expect dramatic effects,
unless one could experimentally break the
time-reversal symmetry suddenly (e.g., by application
of a magnetic field), or could selectively block
the current injected in some of the six
equivalent ellipsoids.  In those cases,
a sudden jump between a semiclassical
regime and a quantum one
should be observable.
It should be noted that band structure effects results 
in k-space current with enough $k_{\parallel}$
to have the electrons injected 
into the
outer conduction-band minima of Si
(the central one is forbidden because of the gap
in that direction),
and explains the long standing puzzle
of why the threshold on Au/Si(111) and
Au/Si(100) is nearly the same:
our calculations show how
the similar results obtained for both  
interfaces are related to these nontrivial distributions
in k-space,
after the appropriate folding
of the gold Brillouin zone inside the
silicon one is performed\cite{applied}.

Next, we compute theoretical I(V) curves from Eq.~(2).
A quantitative comparison
with BEES experiments\cite{bell111},
will then allow us to discuss also the electronic
mean free paths.
First of all, we try the hypothesis
of ballistic electrons.
On intuitive grounds this should suffice for
low temperature, low voltages, and very
thin layers. 
In Fig. 2 we compare experimental
results for 
Au/Si(111) at T=$77$ K, d=$75$ {\AA}\cite{bell111} with
a pure ballistic calculation ($\eta$ very
small and injection at first attempt).
It is clear from these
results that, {\em without using any adjustable
parameter}, the onset is 
reasonably explained
by a purely ballistic theory
that {\em uses 
the right current distributions in k-space}.
Therefore, we are able to give a reasonable
explanation of the experiment for voltages
near the threshold, but
it is also noticed in
Fig. (2) that
data beyond V=$1.2$ eV can 
only be consistently interpreted by assuming an attenuated wave.
To introduce an attenuation mechanism, people 
have considered three major sources of damping:
electron-electron, electron-phonon and electron-defect
interaction.
As the electron-phonon contributions are
greatly reduced at $77$ K we
first consider a $\lambda_{att}(E)$ dominated by 
the electron-electron interaction.
Within an RPA approximation for a
free electron gas with a density
representing gold ($r_{s}=3.01$), we obtain:

\begin{equation}
\lambda_{att}(E) = \lambda^{0}
{ E/E_{F} \over (E-E_{F})^2 }
\end{equation}
\noindent
with $\lambda^{0} =260$ {\AA} (eV)$^{2}$.
Results considering multiple reflections\cite{bell111}
between the surface and the interface through
a {\it specular} model are presented in Fig. (2),
where an excellent agreement is seen again
up to $1.2$ eV. Beyond that voltage, a reduction
in the attenuation length
by about 20\% on average is required to
bring experimental and theoretical intensities
close together. The resulting $\lambda_{att}(E)$
is displayed in the inset of Fig. (2). The 
reduction with respect to the first-principles
RPA approximation 
might be understood as representing
either band-structure or impurity effects
in the effective electron-electron interaction.
It is remarkable the good agreement obtained
for low voltages, where the $\lambda_{att}$
changes quickly with energy following
an RPA-like behaviour. This is at variance 
with E-space Monte-Carlo simulations whereby 
a smoother dependence of  $\lambda_{att}$ with 
energy was found \cite{bell111}. Our results 
suggest, however, that for thin films and 
low voltages, the main source of damping 
is the electron-electron interaction that is 
well described within a RPA approach.

\vskip -5cm

\begin{figure}
\epsfig{figure=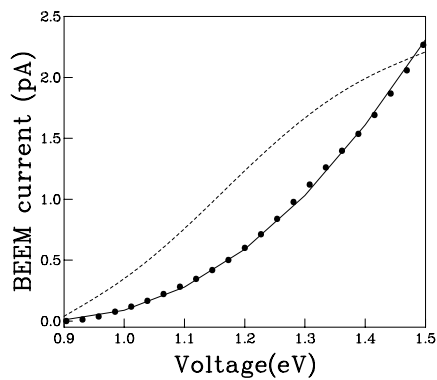,width=8cm}
\caption{
Theoretical I(V) curves for Au/Si(111), 
d=300 \AA\ (experimental values 
-solid circles-
from
ref. 17): 
RPA with $\lambda^{0}=175$ {\AA}(eV)$^{2}$
-dashed line-,
and 
with $\lambda_{att}=125$ {\AA} 
for all E
-solid line-.
}
\end{figure}

However, a different example of BEES-data, where a 
pure ballistic theory is not sufficient even near 
the threshold, is afforded by the case of thick 
layers (see Fig. 3).
In this case we notice that if we use a 
RPA-like energy dependence for
$\lambda_{att}$,   
we find both a discrepancy
in magnitude and a different voltage dependence
for I(V) (as seen in the different slopes). If we
choose a different $\lambda^{0}$
in the RPA expression to get the right
magnitude, we still would observe a serious
discrepancy with the experiment (e.g., see
Fig. (3) where $\lambda^{0}$ has been
reduced to $175$ {\AA} (eV)$^{2}$).
Because all the other elements in the
theory that might be responsible for
the discrepancy ($J(E)$ and $T(E)$ in formula (4))
are calculated from first principles,
we take this as a serious indication
of a different dependence of $\lambda_{att}(E)$
with E.
A possible physical origin for this effect is
the likely presence of
defects (e.g. vacancies)\cite{ventrice}. 
The natural choice for this scenario is
an energy-independent attenuation
length in the Green function.
With this assumption we obtain an excellent agreement with the
experiment ($T=77$ K, $d=300$ {\AA}\cite{bell111})
for $\lambda_{att}=125$ {\AA},
as seen by the solid line in Fig.~3.
This value is in reasonable agreement
with attenuation lengths derived by
different groups in films of similar
thickness\cite{bell111,ventrice} and 
suggests a different behaviour of $\lambda_{att}(E)$
between thick ($300$ {\AA}) and thin ($75$ {\AA}) films.

In conclusion, we have introduced a
Green's function formalism that in
the ballistic limit is an {\it ab initio}
approach to BEEM. 
The particular k-space current distributions
determined by band structure effects 
are the main result of our analysis and
crucial for a quantitative comparison with experimental
BEEM data.
Inelastic effects have also been approximately included by
use of an imaginary self-energy.
This single quantity is fitted to the experiments
to explain a number of spectroscopic data
on the Au/Si interface.

We acknowledge financial support from 
the Spanish CICYT under
contracts number PBB94-53 and PB92-0168C.
K.R. is grateful for financial support from
SFB292 (Germany). We are grateful to Prof. 
K. Heinz for his support.

\end{document}